\documentstyle[sprocl]{article}

\def\Journal#1#2#3#4{{#1} {\bf #2}, #3 (#4)}

\def\be{\begin{equation}}
\def\ee{\end{equation}}
\def\ba{\begin{eqnarray}}
\def\ea{\end{eqnarray}}

\begin{document}

\title{BINARY NONLINEARIZATION OF LAX PAIRS}
\author{W.X. MA
\footnote{\ \ On leave of absence from Institute of 
Mathematics, Fudan University, Shanghai 200433, China} \ 
and B. FUCHSSTEINER}

\address{FB17, Mathematik-Informatik, Universitaet Paderborn,
\\ D-33098 Paderborn, Germany }

\maketitle\abstracts{A kind of Bargmann
 symmetry constraints involved in Lax pairs and adjoint Lax pairs
is proposed for soliton hierarchy.
The Lax pairs and adjoint Lax pairs are nonlinearized into
a hierarchy  of  commutative finite dimensional integrable Hamiltonian
systems and explicit integrals of motion may also be generated.
The corresponding 
binary nonlinearization procedure leads to  
 a sort of  involutive  solutions
to every system in soliton hierarchy which are all of finite gap.
 An illustrative example
is given in the case of AKNS soliton hierarchy.}

\section{Introduction}

Symmetry 
constraints become prominent because of the 
important role 
they play in the soliton theory \cite{AntonowiczWojciechowski} 
\cite{KonopelchenkoStrampp}
\cite{OevelStrampp} \cite{RagniscoW}. 
A kind of very successful symmetry
constraint method for soliton equations is proposed through the
nonlinearization technique called mono-nonlinearization \cite{Cao} 
\cite{GengMa}.
However, mono-nonlinearization  involves only the Lax
pairs of soliton equations. 
We would like to elucidate that the 
mono-nonlinearization technique
can successfully be extended to the Lax
pairs and the adjoint Lax pairs associated with
soliton hierarchy. 
The corresponding symmetry constraint procedure is
called a binary nonlinearization technique \cite{Ma} \cite{MaStrampp} 
\cite{Geng} because it involves the
Lax pairs and the adjoint Lax pairs and puts the linear Lax pairs
into the nonlinearized Lax systems. A kind of useful symmetries in our
symmetry constraints is exactly the
specific symmetries expressed through the variational derivatives of the 
potentials.
The resulting theory 
provides a
method of separation of variables for solving nonlinear soliton
equations 
and exhibits integrability by
quadratures for soliton equations.
It also narrows
the gap between finite dimensional integrable Hamiltonian systems and
infinite dimensional integrable soliton equations.
An illustrative example is
carried out in the case of the 
 three-by-three matrix spectral problem for
AKNS soliton hierarchy. 

\section{Basic idea of binary nonlinearization}

This section reveals how to manipulate a binary nonlinearization procedure
for a given soliton hierarchy along with a basic idea for the proof of the 
main result.
Let $\cal B$ denote the differential algebra of differential vector functions
$u=u(x,t)$, and write for $k\ge 0$
\[  
{\cal V}^s_{(k)} 
=\{ (P^{ij}\partial  ^k )_{s\times s}
\left.\right | P^{ij}\in 
{\cal B} \},\ 
\widetilde {\cal V}^s_{(k)}= 
{\cal V}^s_{(k
)}\otimes
C[\lambda,\lambda^{-1}],\ \partial  =\frac d{dx} .\]
For 
$U=U(u,\lambda  )\in
\widetilde{\cal V}_{(0)}^s$, we choose a solution 
to the adjoint representation equation $V_x=[U,V]$:
\[
V=V(u,\lambda  )=\sum _{i\ge0}V_i\lambda  ^{-i}, \  V_i\in 
{\cal V}^s_{(0)}.\] 
Suppose that 
the isospectral ($\lambda  _{t_n}=0$) compatibility conditions
$U_{t_n}-V^{(n)}_x+[U,V^{(n)}]=0,\ n\ge0,$
of the Lax pairs 
\[ \left \{ \begin{array}{l}\phi _x=U\phi =U(u,\lambda  )\phi ,\ U\in
\widetilde {\cal V}_{(0)}^s \vspace {2mm}
\\ 
\phi _{t_n}=
V^{(n)}
\phi =
V^{(n)}
(u,\lambda  )\phi ,\ 
V^{(n)}=(\lambda  ^{n}V)_++\Delta _n,\ 
\Delta _n\in \widetilde {\cal V}_{(0)}^s\end{array}
\right.\]
determine a soliton hierarchy 
\begin{equation}  u_{t_n}=K_n=JG_n=J\frac {\delta H_n}{\delta u},\ n\ge0. 
\end{equation} 
If 
$\phi =(\phi _1,\phi_2,\cdots, \phi_s)^T$ and $\psi =(\psi _1,\psi_2,
\cdots, \psi_s)^T$ satisfy the spectral problem and the adjoint spectral
problem
\[ \phi _x=U(u,\lambda  )\phi,\ \psi _x=-U^T(u,\lambda  )\psi,\]
and we set the matrix 
$\bar V =\phi \psi ^T= (\phi _k\psi _l)_{s\times s }$,
then we have the following two basic results used in binary nonlinearization
\cite{FokasAnderson} \cite{MaStrampp}:

\noindent (i) the variational derivative of the spectral parameter $\lambda  $
with respect to the potential $u$ may be expressed by 
\be
 \frac {\delta \lambda  }{\delta u}=
\frac {\textrm{tr}\bigl(\bar V\frac{\partial  U}{\partial  u}\bigr)}{
-\int _{-\infty}^\infty \textrm{tr}\bigl(\bar V\frac {\partial  U}
{\partial  \lambda 
}\bigr) dx},\label{vdoflambda}
\ee

\noindent (ii) the matrix 
$\bar V $
is a solution to the adjoint representation equation
 $ V _x=[U, V]$, i.e. $\bar V _x=[U,\bar V]$.

Now introduce distinct eigenvalues $\lambda  _1,\cdots, \lambda  _N$ and let
\[\phi^{(j)}=(\phi_{1j},\cdots,\phi_{sj})^T,
\ \psi^{(j)}=(\psi_{1j},\cdots,\psi_{sj})^T\ \, (1\le j\le N) \]
 denote the eigenvectors and the 
adjoint eigenvectors corresponding to $\lambda  _j\ (1\le j\le N)$, 
respectively.
Make the Bargmann {\bf  symmetry constraint}
\be 
K_0=JG_0=J\sum_{j=1}^N E_j \frac {\delta \lambda  _j }
{\delta u}\ \ \textrm{or} \ \ \ 
  G_0=\sum_{j=1}^N E_j \frac {\delta \lambda  _j }{\delta u},
\ee 
where $E_j=
-\int _{-\infty}^\infty <\bar V(\lambda  _j),\frac {\partial  U}
{\partial  \lambda  _j
}>dx$, $\bar V(\lambda  _j)=\phi^{(j)}\psi^{(j)T},$ $1\le j\le N$. 
The Bargmann constraint requires the covariant $G_0$ to be a
potential function not including any potential differential
and hence from the Bargmann symmetry 
constraint we may find 
 an explicit nonlinear expression for the potential
\be  u=f
(\phi ^{(1)},\phi ^{(2)},\cdots,\phi ^{(N)};\psi^{(1)},\psi^{(2)},
\cdots,\psi^{(N)})
.\label{u}\ee 
Upon instituting (\ref{u}) into the Lax pairs and the
adjoint Lax pairs, we get
two nonlinearized Lax systems, i.e. the nonlinearized spatial system 
\be \left \{ \begin{array}{l}  
\phi _{jx}=U(f
(\phi ^{(1)},\cdots,\phi ^{(N)};\psi^{(1)},\cdots,\psi^{(N)})
,\lambda  _j)
\phi _{j},\ 1\le j\le N,\vspace{2mm}\\ 
\psi _{jx}=-U^T(f
(\phi ^{(1)},\cdots,\phi ^{(N)};\psi^{(1)},\cdots,\psi^{(N)})
,\lambda  _j)
\psi_{j},\ 1\le j\le N;\end{array}\right.  \label{nxpart}\ee 
and the nonlinearized temporal systems for $n\ge 0$
\be \left \{\begin{array}{l}
\phi _{jt_n}=V^{(n)}(f
(\phi ^{(1)},\cdots,\phi ^{(N)};\psi^{(1)},\cdots,\psi^{(N)})
,\lambda  _j)
\phi_{j},\ 1\le j\le N,\vspace{2mm}\\ 
\psi _{jt_n}=-V^{(n)T}(f
(\phi ^{(1)},\cdots,\phi ^{(N)};\psi^{(1)},\cdots,\psi^{(N)})
,\lambda  _j)
\psi_{j},\ 1\le j\le N.\end{array} \right.\ \label{ntpartn} \ee 

In order to discuss the integrability of (\ref{nxpart}) and (\ref{ntpartn}),
we choose the symplectic structure $\omega ^2$  on ${\mbox{\rm I 
\hspace{-0.9em} R}}  ^{2sN}$
\[ \omega ^2=\sum _{i=0}^s\sum _{j=0}^N  d\phi_{ij}\wedge d\psi_{ij}=
\sum _{i=0}^sdP_i\wedge dQ_i,\]
where $P_i=(\phi_{i1},\cdots,\phi_{iN})^T,\ 
Q_i=(\psi_{i1},\cdots,\psi_{iN})^T$, $1\le i\le s$.
We accept  the following corresponding Poisson bracket  
for two functions $F,G$ defined over the phase space  ${\mbox{\rm I 
\hspace{-0.9em} R}}  ^{2sN}$
\begin{eqnarray}&&\{F,G\}=\omega ^2(IdG,IdF)=\omega ^2(X_G,X_F)\nonumber
\\&&=
\sum_{i=1}^s(<\frac {\partial  F}{\partial  Q_i},\frac {\partial  G}
{\partial  P_i}>-
<\frac {\partial  F}{\partial  P_i},\frac {\partial  G}{\partial  Q_i}>),
\label{poissonbracket}
\end{eqnarray}
where $IdH=X_H$ represents the Hamiltonian vector field with energy
 $H$ defined by
$i_{IdH}\omega ^2=i_{X_H}\omega ^2=dH$
 and $<\cdot,\cdot>$ represents the standard inner product of ${\mbox{\rm I \hspace{-0.9em} R}}  ^N$.
Then we  accept the following
corresponding Hamiltonian system with the Hamiltonian function $H$
\be 
\dot {P}_i=\{P_i,H\}=-\frac {\partial  H}{\partial  Q_i},\ 
\dot {Q}_i=\{Q_i,H\}=\frac {\partial  H}{\partial  P_i},\ 1\le i\le s.\ee 

\noindent {\bf Main Result:} {\it The nonlinearized spatial system 
(\ref{nxpart})
  is a finite
dimensional integrable Hamiltonian system in the Liouville sense,
 and the 
nonlinearized temporal systems (\ref{ntpartn}) for $n\ge0$
 may be transformed into  a hierarchy of finite dimensional 
integrable Hamiltonian systems in the Liouville sense, 
under the control of 
the nonlinearized spatial system (\ref{nxpart}).
Moreover the potential $u=f$
 determined by the Bargmann symmetry constraint
solves the $n$-th soliton equation $u_{t_n}=K_n$ in the hierarchy.}

\noindent {\bf Idea of Proof:} 
Note that we have 
\[(V(f,\lambda  ))_{x}=[U(f,\lambda  ), V(f,\lambda  )],
\ (\bar V(\lambda  _j))_{x}=[U(f,\lambda  _j),\bar V(\lambda  _j)]\]
and when $ u_{t_n}=K_n,$ we have
  \[ ( V(f,\lambda  ))_{t_n}=[V^{(n)}(f,\lambda  ), V(f,\lambda  )],\  
(\bar V(\lambda  _j))_{t_n}=[V^{(n)}(f,\lambda  _j),\bar V(\lambda  _j)].\]
Therefore we may show that $F=\frac12 \textrm{tr}(V(f,\lambda  ))^2$ 
is a common generating function for
integrals of motion of (\ref{nxpart}) and (\ref{ntpartn}) since 
$F_x=\frac12 \textrm{tr}(V^2)_x
=\frac12\textrm{tr} [U,V^2]=0$ and 
$F_{t_n}=\frac12 \textrm{tr}(V^2)_{t_n}
=\frac12 \textrm{tr}[V^{(n)},V^2]=0$. A similar deduction may verify that 
$\bar F_j=\frac12 \textrm{tr}(\bar V(\lambda  _j))^2$, $1\le j\le N$, are 
integrals of motion of (\ref{nxpart}) and (\ref{ntpartn}), too.
Noticing 
\be F=\sum_{n\ge0}F_n\lambda  ^{-n},\ \bar F_j=\frac12 (\sum_{i=1}^s
\phi_{ij}\psi_{ij})^2,\ 
1\le j\le N,\ee 
 we get a series of explicit
integrals of motion: 
$\bar F_j,\ 1\le j\le N,\ \{F_n\}_{n=0}^\infty,$
which may be proved to be involutive with respect to the Poisson
bracket (\ref{poissonbracket}). Further it is not difficult to show the 
Liouville integrability of  (\ref{nxpart}) and (\ref{ntpartn}) when
they can be rewritten as Hamiltonian systems with Hamiltonian functions 
being polynomials in $F_m,\ m\ge1$.

In addition, because the compatibility condition
of (\ref{nxpart}) and (\ref{ntpartn}) is still the $n$-th soliton equation
$u_{t_n}=K_n$,
$u=f(\phi_j;\psi_j)$ gives an involutive solution
to the $n$-th soliton equation $u_{t_n}=K_n$ once 
$\phi_j,\,\psi_j,\ 1\le j\le N,$ solve
(\ref{nxpart}) and (\ref{ntpartn}),
simultaneously. This sort of involutive solutions 
also exhibits a kind of separation of independent variables $x, t_n$ for
soliton equations.

\section{The case of AKNS Hierarchy}
For AKNS hierarchy, we introduce a three-by-three matrix spectral problem
\[ 
\left(\begin{array}{c}
 \phi  _1 \\ \phi _2 \\ \phi _3\end{array}\right) _x=U\left(\begin{array}{c}
 \phi _1 \\ \phi _2 \\ \phi _3\end{array}\right) =
\left(\begin{array}{ccc}
 -2\lambda  & \sqrt{2}q&0 \vspace{1mm}\\ \sqrt{2}r&0 &\sqrt{2}q\vspace{1mm}
\\ 0&\sqrt{2}r &2\lambda  
\end{array}\right)\left(\begin{array}{c}
 \phi _1 \\ \phi _2 \\ \phi _3\end{array}\right).\]
In this case, 
 $ \phi =(
 \phi _1 , \phi _2 , \phi _3)^T $  and $ u =
( q  , r)^T$.
A hierarchy of AKNS soliton equations \cite{MaStrampp}
\be  u_{t_n}=
K_n=
\left(\begin{array}{c} -2b_{n+1}\\2c_{n+1}\end{array}\right) =
JL^n
\left(\begin{array}{c} r\\q\end{array}\right)=J\frac {\delta H_n}{\delta u}
,\ n\ge 0\ee 
is the compatibility
conditions of the Lax pairs
\be  \phi _x=U\phi ,\ \phi _t=V^{(n)}\phi ,\ V^{(n)}=(\lambda  ^nV)_+\,.
\ee 
Here the operator solution $V$ to
$V_x=[U,V]$,
the Hamiltonian operator $J$, the recursion operator $L$, and the
Hamiltonian functions $H_n$ for $n\ge0$ read as
\[
V=\left(\begin{array}{ccc} 2a&\sqrt{2}b&0 \vspace{1mm}\\ \sqrt{2}c&0&\sqrt{2}b
\vspace{1mm}\\
0&\sqrt{2}c&-2a\end{array}\right) 
=\sum _{i=0}^\infty
\left(\begin{array}{ccc} 2a_i&\sqrt{2}b_i&0\vspace{1mm}\\
\sqrt{2}c_i&0&\sqrt{2}b_i\vspace{1mm}\\
0&\sqrt{2}c_i&-2a_i\end{array}\right) \lambda  ^{-i},\]
\[
J=
\left(\begin{array}{cc} 0&-2\vspace{1mm}\\2&0\end{array}\right),\ 
L=
\left(\begin{array}{cc} \frac12 \partial  -r\partial  ^{-1}q&r\partial  ^{-1}r
\vspace{1mm}\\
-q\partial  ^{-1}q&-\frac12 \partial  +q\partial  ^{-1}r\end{array}\right) ,\ 
H_n=\frac {2a_{n+2}}{n+1}.\]
The operators $J$ and $JL$ constitute a Hamiltonian pair and 
$L^*$ is hereditary \cite{Fuchssteiner1}. 

In this AKNS case, the Bargmann symmetry constraint becomes
\be  
K_0=J\frac {\delta H_0 }{\delta u}
=J\sum 
_{j=1}^N \left(\begin{array}{c} \sqrt{2}(\phi _{2j}\psi _{1j}+
\phi_{3j}\psi_{2j})\\ \sqrt{2}(\phi _{1j}\psi_{2j}+
\phi_{2j}\psi_{3j})\end{array}\right),\ee 
which engenders an explicit expression for the potential $u$
\be u
=f(\phi_{ij};\psi_{ij})=\sqrt{2}\left(\begin{array}{c} 
< P  _1,Q_2>+< P  _2,Q_3>\\
< P  _2,Q  _1>+< P  _3,Q  _2>
\end{array}\right).\label{uakns}\ee  
Further besides $\bar F_j$, $1\le j\le N$, 
we can directly give the following   explicit integrals
of motion for the nonlinearized Lax systems
 \[ \begin{array}{l}
F:=\frac12\textrm{tr}V^2=4(a^2+bc)=\sum_{m\ge0}F_m\lambda  ^{-m},\vspace{2mm}\\ 
F_0=4,\ F_1=-8(<P_1 ,Q_1>-<P_3,Q_3>),\vspace{2mm}\\
F_m=4\sum _{i=1}^{m-1}\bigl[
(<A^{i-1}P _1,Q _1>-<A^{i-1}P_3,Q_3>)\times \vspace{2mm}\\
\qquad (<A^{m-i-1}P _1,Q _1>-<A^{m-i-1}P_3,Q_3>)\vspace{2mm}
\\ \qquad 
+ 2(<A^{i-1}P_1,Q_2>+<A^{i-1}P_2,Q_3>)\times \vspace{2mm}\\
\qquad (<A^{m-i-1}P_2,Q_1>+<A^{m-i-1}
P_3,Q_2>)\bigr]\vspace{2mm}\\ \qquad
-8<A^{m-1}P _1,Q _1>-<A^{m-1}P_3,Q_3>,
\ m\ge 2,\end{array}\]
where $A=\textrm{diag}(\lambda  _1,\lambda  _2,\cdots,\lambda  _N)$.
The nonlinearized 
spatial system (\ref{nxpart}) is rewritten as an integrable Hamiltonian system
\be  P  _{ix}=\{P_i,H\}=-\frac {\partial  H}{\partial   Q  _{i}},\ 
 Q  _{ix}=\{Q_i,H\}=\frac {\partial  H}{\partial   Q  _i},\ 
i=1,2,3\ee 
with the Hamiltonian function
\[\begin{array}{l}H=2(<A P _1,Q  _1>-<AP _3,Q  _3>)\vspace{2mm}\\
-2(<P _1,Q  _2>+<P _2,Q  _3>)(<P _2,Q  _1>+<P _3,Q  _2>)
,\end{array}\]
and under the control of the nonlinearized spatial system (\ref{nxpart}),
the nonlinearized temporal systems (\ref{ntpartn}) for $n\ge 0$ can also be
rewritten as the integrable Hamiltonian systems
\be  P  _{it_n}=\{P_i,H_n\}=-\frac {\partial  H_{n}}{\partial   Q  _i},\ 
 Q  _{it_n}=\{Q_i,H_n\}=\frac {\partial  H_{n}}{\partial  P  _i},\ 
\ i=1,2,3\ee 
with the Hamiltonian functions
\[H_n= -\frac14
\sum _{m=0}^n \frac {d_m}{m+1}
\sum_{\begin{array}{c}{\scriptstyle{ i_1+\cdots+i_{m+1}=n+1}}
\vspace{-1mm}
\\ {\scriptstyle{ i_1,\cdots,i_{m+1}\ge1}}\end{array}}
F_{i_1}
\cdots F_{i_{m+1}}\, ,\]
where the constants $d_m$ are defined by 
\[ \begin{array}{l}d_0=1,\ d_1=-\frac 18,\ d_2=\frac3{128},\vspace{2mm}\\
 d_m=-\frac12
\sum_{i=1}^{m-1}d_id_{m-i}-\frac14d_{m-1}-\frac18 
\sum_{i=1}^{m-2}d_id_{m-i-1},\
m\ge3.\end{array}\]
Moreover following the previous main result, the potential (\ref{uakns}) 
with 
\[P_i(x,t_n)=g^x_{H}g^{t_n}_{H_n}P_i(0,0),\  
Q_i(x,t_n)=g^x_{H}g^{t_n}_{H_n}Q_i(0,0),\  i=1,2,3,\]
gives rise to a sort of involutive solutions
with separated variables $x, t_n$ to 
the $n$-th AKNS soliton equation $u_{t_n}=K_n$.
Here $g^y_G$ denotes the Hamiltonian phase flow of $G$ with a parameter 
variable $y$ but $P_i(0,0),\,Q_i(0,0)$ may be arbitrary initial value vectors.
A finite gap property for the resulting involutive solutions may also be
shown. 

\section{Concluding remarks}

We remark that the finite dimensional Hamiltonian systems
generated by nonlinearization technique depend
on the starting Lax pairs.
Thus the same equation may be connected with different 
finite dimensional Hamiltonian systems once it possesses different 
Lax pairs. AKNS soliton equations are exactly such examples \cite{MaStrampp}.

We  also point out that 
the Neumann symmetry constraint
and the higher order symmetry constraints
\be K_{-1}=J\sum _{j=1}^NE_j\frac {\delta \lambda  _j}{\delta u},\ 
K_m=JG_m=J\sum _{j=1}^NE_j\frac {\delta \lambda  _j}{\delta u},\ (m\ge1),\ee 
may be considered.
These two sorts  of symmetry constraints are somewhat different from the
Bargmann symmetry constraints because $K_{-1}$ is a constant vector
and the conserved covariants 
$G_m,\, m\ge1, $ involve some differentials of the potential.
This suggests that a few new tools are needed for 
 discussing  them \cite{Zeng}.
Similarly, we can consider the corresponding
 $\tau$-symmetry (i.e. time first order  dependent 
symmetry \cite{Chen}) constraints or more generally,
time polynomial dependent symmetry \cite{Fuchssteiner2} constraints . 
Binary nonlinearization may also be well applied  to discrete systems
and non-Hamiltonian soliton equations such as the Toda lattice and the coupled 
Burgers equations \cite{Ma0}.
Note that in the case of KP hierarchy, the similar Bargmann 
symmetry constraints have been carefully analyzed as well \cite{OevelStrampp},
and the specific symmetries we use in constraints
are sometimes called additional symmetries \cite{Dickey}
and are often taken as
source terms of soliton equations
\cite{Mel'nikov}. It should also be noted that 
the nonliearized Lax systems are intimately related to 
stationary equations \cite{Tondo} and the more general 
nonliearized Lax systems can be generated from the linear combination of 
Bargmann symmetry constraints which will be shown in a late publication.

However, in the binary nonlinearization procedure, there exist two
intriguing {\bf open problems}. The first one  is 
why the nonlinearized spatial system (\ref{nxpart})
and the nonlinearized temporal systems (\ref{ntpartn}) for $n\ge 0$
with the control of the nonlinearized 
spatial system (\ref{nxpart}) always possess
 Hamiltonian structures? The second one is 
whether or not the nonlinearized temporal systems (\ref{ntpartn})
for $n\ge 0$ 
are themselves integrable soliton equations 
without the control of the nonlinearized spatial system (\ref{nxpart}). 
These two problems are important and interesting  
but need some further investigation.

\section*{Acknowledgments}One of the authors (W. X. Ma) 
would like to thank the 
Alexander von Humboldt Foundation for a research fellow award and 
the National Natural 
Science Foundation of China and the Shanghai Science and Technology Commission
of China for their financial support. 
He is also grateful to Drs. W. Oevel, P. Zimmermann
and G. Oevel for their helpful and stimulating discussions.


\section*{References}
\begin{thebibliography}{99}
\bibitem{AntonowiczWojciechowski}M. Antonowicz and S. Wojciechowski,
{\em J. Math. Phys.} {\bf 33}, 2115 (1992).
\bibitem{Cao}C.W. Cao, 
{\em Sci. China} A {\bf 33}, 528 (1990). 
\bibitem{Chen}H.H. Chen, Y.C. Lee and J.E. Lin, in {\em Advances in 
Nonlinear Waves}, Vol.2, ed. L. Debnath (Pitman, New York, 1985), p233.
\bibitem{Dickey}L.A. Dickey, 
{\em Lett. Math. Phys.} {\bf 34}, 379 (1995).
\bibitem{FokasAnderson}A.S. Fokas and R.L. Anderson,
{\em J. Math. Phys.} {\bf 23}, 1066 (1982).
\bibitem{Fuchssteiner1}B. Fuchssteiner, \Journal{{\it 
Nonlinear Anal. Theor. Meth. Appl.}}{3}{849}{1979}.
\bibitem{Fuchssteiner2}B. Fuchssteiner, \Journal{{\it 
Prog. Theor. Phys.}}{70}{1508}{1983}.
\bibitem{Geng}X.G. Geng, 
{\em Phys. Lett.} A {\bf 194}, 44 (1994).
\bibitem{GengMa}X.G. Geng and W. X. Ma, {\em Nuovo Cimento} A {\bf 108},
477 (1995). 
\bibitem{KonopelchenkoStrampp}B. Konopelchenko and W. Strampp,
{\em J. Math. Phys.} {\bf 33}, 3676 (1992).
\bibitem{Ma0} W.X. Ma, {\em J. Phys. A: Math. Gen.} {\bf 26}, L1169 (1993).
\bibitem{Ma} W.X. Ma, 
{\em J. Phys. Soc. Jpn.} {\bf 64}, 1085 (1995);
Symmetry constraint of MKdV equations by binary nonlinearization,
to appear in {\em Physica} A.
\bibitem{MaStrampp}W.X. Ma and W. Strampp, 
{\em Phys. Lett.} A {\bf 185}, 277 (1994).
\bibitem{Mel'nikov} V.K. Mel'nikov,
{\em J. Math. Phys.} {\bf 31}, 1106 (1990).
\bibitem{OevelStrampp}W. Oevel and W. Strampp,
{\em Commun. Math. Phys.} {\bf 157}, 51 (1993).
\bibitem{RagniscoW}O. Ragnisco and S. Wojciechowski, 
{\em Inverse Problems} {\bf
8}, 245 (1992).
\bibitem{Tondo}G. Tondo, {\em Theor. Math. Phys.} {\bf 33}, 796 (1994).
\bibitem{Zeng}Y.B. Zeng,
{\em Physica} D {\bf 73}, 171 (1994).
\end{thebibliography}
\end{document}